# A Study of the Speedups and Competitiveness of FPGA Soft Processor Cores using Dynamic Hardware/Software Partitioning

Roman Lysecky and Frank Vahid*
Department of Computer Science and Engineering
University of California, Riverside
{rlysecky, vahid}@cs.ucr.edu, http://www.cs.ucr.edu/{~rlysecky,~vahid}
*Also with the Center for Embedded Computer Systems at UC Irvine

**Abstract**

*Field programmable gate arrays (FPGAs) provide designers with the ability to quickly create hardware circuits. Increases in FPGA configurable logic capacity and decreasing FPGA costs have enabled designers to more readily incorporate FPGAs in their designs. FPGA vendors have begun providing configurable soft processor cores that can be synthesized onto their FPGA products. While FPGAs with soft processor cores provide designers with increased flexibility, such processors typically have degraded performance and energy consumption compared to hard-core processors. Previously, we proposed warp processing, a technique capable of optimizing a software application by dynamically and transparently re-implementing critical software kernels as custom circuits in on-chip configurable logic. In this paper, we study the potential of a MicroBlaze soft-core based warp processing system to eliminate the performance and energy overhead of a soft-core processor compared to a hard-core processor. We demonstrate that the soft-core based warp processor achieves average speedups of 5.8 and energy reductions of 57% compared to the soft core alone. Our data shows that a soft-core based warp processor yields performance and energy consumption competitive with existing hard-core processors, thus expanding the usefulness of soft processor cores on FPGAs to a broader range of applications.*

## 1. Introduction

Field programmable gate arrays (FPGAs) provide great flexibility to hardware designers. While past designers primarily used FPGAs for prototyping and debugging purposes, many commercial end-products now incorporate FPGAs. Designers using FPGAs can quickly build entire systems or hardware components while still leveraging the prototyping and debugging advantages that FPGAs have over ASIC (application-specific integrated circuit) designs. Continuing increases in FPGA capacity, performance, and architectural features are enabling more designs to be implemented using FPGAs. Additionally, FPGAs costs are decreasing, allowing designers to incorporate FPGAs with one million equivalent gates for less than $12 [27].

While designers can use FPGAs to quickly create efficient hardware designs, many systems require a combination of both software and hardware. In the late 1990s, FPGA vendors began introducing single-chip microprocessor/FPGA devices. Such devices include one or more hard-core (implemented directly using IC transistors/gates) microprocessors and an FPGA fabric on a single IC, and provide efficient mechanisms for communication between the microprocessor and FPGA. Atmel [2] and Triscend [23] were the first to make these devices available, both incorporating low-end microprocessors and FPGAs supporting tens of thousands of gates. More recently, Altera developed the Excalibur devices having an ARM9 processor and a one million gate FPGA [1]. Xilinx offers the VirtexII Pro devices incorporating two or more PowerPC processors and an FPGA fabric with tens of millions of gates [26].

While single-chip hard-core microprocessor/FPGA platforms offer excellent packaging and communication advantages, a soft-core approach offers the advantage of flexibility and lower part costs. Many FPGA vendors are now offering such soft processor cores that designers can implement using a standard FPGA. Altera offers both the NIOS and more recently the NIOS II soft processor cores [1]. The NIOS II processor is a 32-bit configurable processor supporting clock frequencies as high as 135 MHz. Xilinx offers the PicoBlaze and MicroBlaze soft processor cores [26]. The MicroBlaze processor is 32-bit configurable processor core capable of supporting clock frequencies as high 150 MHz. These soft processor cores offer designers tremendous flexibility during the design process, allowing the designers to configure the processor to meets the needs of their systems (e.g., adding custom instructions or including/excluding particular datapath coprocessors) and to quickly integrate the processor within any FPGA. Unlike single-chip microprocessor/FPGA systems using hard-core processors, soft processor cores allow designers to incorporate varying numbers of processors within a single FPGA design depending on an application's needs. While some embedded system designs may require a few processors, other designs can include 64 processors [11] or more. Furthermore, as reported in [11], customers of Tensilica [22], who provide customizable soft processor core solutions, are creating chip designs that incorporate over five processors on average.

Unfortunately, soft processor cores implemented using FPGAs typically have higher power consumption and decreased performance compared with hard-core processors. To alleviate the performance and power overhead, a designer can potentially use hardware/software partitioning to increase software performance while decreasing energy. Hardware/software partitioning is the process of dividing an application among software (running on a microprocessor) and hardware co-processors. By identifying the critical kernels within the software application, one can re-implement those software kernels as a hardware coprocessor on the FPGA. Extensive research has shown that hardware/software partitioning can result in overall software speedups of 200%-1000% [3][7][8][9][14][24], as well as reducing system energy by up to 99% [12][13][20][25].

However, hardware/software partitioning requires a special compiler that profiles, estimates hardware size, and generates an application binary that communicates with a hardware description that implements the software kernels. Thus, partitioning imposes a significant increase in tool complexity, and results in non-standard output having greatly reduced portability compared to a standard binary. Recently, we showed [21] that designers could perform desktop hardware/software partitioning starting from binaries rather than from high-level code, with competitive resulting performance and energy. Binary-level partitioning approaches can produce excellent results by using decompilation techniques to retrieve most of the high-level information typically lost at the binary level [5].

Binary-level partitioning opens the door to dynamic hardware/software partitioning, in which an executing binary is dynamically optimized by moving software kernels to configurable logic, a process we call *warp processing* [15][16][19]. However, warp processors previously only targeted single-chip multiprocessor/FPGA devices incorporating a hard-core processor. Extensive details of warp processing are beyond the scope of this paper and appear in other publications. Our purpose in this paper is to study whether warp processing methods could potentially make a soft processor core competitive with a hard-core processor with respect to performance and energy. Because warp processing occurs dynamically and transparently, soft-core warp processing could open the door to a much wider use of FPGA soft-cores.

In this paper, we investigate the benefits of warp processing for soft processor cores. We present a warp processing system consisting of a soft processor core that a designer can implement using any FPGA. While we could potentially target any soft processor core, we focus our efforts on the MicroBlaze processor. By utilizing a warp processor based on a soft processor core, a designer can quickly implement a software system using a low cost FPGA, potentially incorporating several processors, with increased performance and lower energy consumption compared with the soft processor core alone, and comparable with hard-core processors.

## 2. MicroBlaze Soft Processor Core

The MicroBlaze soft processor core provided by Xilinx is a 32-bit configurable processor core. A designer can create a system incorporating a MicroBlaze using the Xilinx Platform Studio in which a designer can quickly build a MicroBlaze processor system by instantiating and configuring cores from the provided libraries. Figure 1 presents a simple MicroBlaze system incorporating the MicroBlaze processor along with several components to create a complete system. The MicroBlaze processor utilizes a three-stage pipeline with variable length instruction latencies typically ranging from one to three cycles. The MicroBlaze has a Harvard memory architecture and utilizes two Local Memory Busses (LMB) for instruction and data memory. The system shown in Figure 1 includes two Block RAMs (BRAM), one for instruction memory and one for data memory, whose sizes are user defined. A local memory bus to BRAM interface connects the MicroBlaze with the instruction and data memories. The system also includes two peripherals connected via the On-Chip Peripheral Bus (OPB). After specifying the system architecture and configuring the MicroBlaze processor, the Xilinx Platform Studio tools synthesize the design and create a bitstream for the system as well a generate a set of software libraries that a design can use to interface with the various components in the system. Finally, a designer can compile their application and combine the application's binary with the bitstream to produce the final system bitstream.

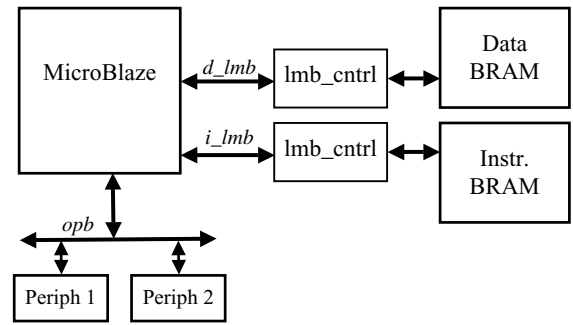

**Figure 1:** Simple MicroBlaze processor system.

Key features of the MicroBlaze processor, as well as other soft processor cores, include the user configurable options that allow a designer to tailor the processor's functionality to their specific design. The MicroBlaze's user-configurable options include configurable instruction and data caches, incorporating a hardware multiplier to enable the *mul* instruction, incorporating a hardware divider to enable the *idiv* instruction, and incorporating a barrel shifter to enable the *bs* and *bsi* instructions.

Many of the MicroBlaze processor's configurable options can have a significant impact on performance. While the impacts on performance of incorporating or excluding caches is well known and widely understood, the impact of other configurable options such as the inclusion of a hardware barrel shifter is also extremely important, especially in embedded systems in which bit manipulation is often used.

We therefore analyzed two embedded system benchmark applications, *brev* and *matmul*, from the Powerstone benchmark suite (which we obtained from Motorola). The critical kernel of the benchmark *brev* performs an efficient bit reversal but heavily relies on shift operations. If the MicroBlaze processor is configured without the hardware barrel shifter or hardware multiplier, the resulting application binary will perform an n-bit shift by using n successive add operations each of which doubles the values of the variable being shifted. Compared with a MicroBlaze processor including a barrel shifter and multiplier, the absence of these configurable options results in a 2.1X longer execution time for the application *brev*. For the application *matmul*, the critical region is a matrix multiplication. Without a hardware multiplier, the compiler will use a software function to perform every multiplication, thereby increasing the execution time for *matmul* by 1.3X. However, with knowledge of the final software application, a designer can reduce the amount of configurable logic used within the FPGA if they do not require a hardware barrel shifter, multiplier, or divider.

Another potential drawback of the MicroBlaze processor is the lack of floating point instructions, requiring software routines to perform these operations. However, as FPGAs continues to increase in complexity, soft processor cores will likely begin to incorporate more functionality possibly allowing designers to configure the processor with a hardware floating point unit.

**Figure 2:** MicroBlaze single-processor warp processing system.

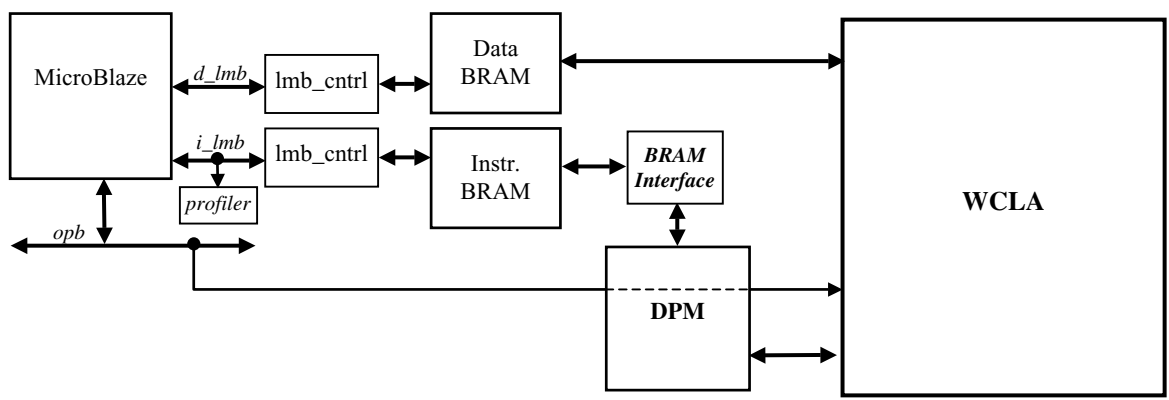

The most significant drawbacks to using a soft core processor are the performance and energy overhead. Although the MicroBlaze processor has a three-stage pipeline, during the execute phase of the pipeline, instructions have different latencies. For example, an addition operation has a latency of only one cycle, whereas a multiply operation requires three cycles. Other instructions such as branch instructions have latencies anywhere from one cycle to three cycles depending on the instruction and whether or not the branch is taken. However, of the applications we analyzed, most branch instructions had a latency of two cycles, as the compiler often did not utilize the branch delay slot. While the performance overhead can significantly impact the overall energy consumption, FPGAs also consume more power than ASICs or custom designs. This increased power consumption and performance overhead results in a system that consumes more energy than a similar hard-core processor.

## 3. MicroBlaze-based Warp Processor

Figure 2 presents a single processor MicroBlaze-based warp processor. The warp processor consists of a main processor with instruction and data caches, an efficient on-chip profiler, a warp configurable logic architecture (WCLA), and a dynamic partitioning module (DPM). Initially, the software application executing on the warp processor will execute only on the MicroBlaze processor. During execution of the application, the profiler monitors the execution behavior to determine the critical kernels within the application. After identifying the critical regions, the dynamic partitioning module re-implements the critical software regions as a custom hardware component within the WCLA using our Riverside On-Chip Partitioning Tools (ROCPART). Extensive discussion of the ROCPART tools is beyond the scope of this paper and has been published in [15][16][19]. We highlight the key features in this section enough to better understand the experimental study that we performed.

The MicroBlaze warp processor includes a *profile*r within the warp processor to determine the critical kernels within the executing application that the warp processor could implement as hardware. Based on the profiler presented in [10], the profiler incorporated in our warp processor design is a non-intrusive profiler that monitors the instruction addresses seen on the local instruction memory bus. Whenever a backward branch occurs, the profiler updates a small cache that stores the branch frequencies.

The dynamic partitioning module executes the ROCPART partitioning and synthesis algorithms. ROCPART first analyzes the profiling results for the executing application and determines which critical region the warp processor should implement in hardware. After selecting the software region to implement in hardware, the DPM accesses the application's software binary by interfacing with the dual ported instruction BRAM. The partitioning tool then decompiles the critical region into a control-dataflow graph and synthesizes the graph to create the hardware circuit [19]. The hardware circuit is then optimized and mapped onto the WCLA by performing technology mapping, placement, and routing to produce the final hardware bitstream [16][17]. Finally, the DPM configures the configurable logic and updates the executing application's binary code to utilize the hardware within the configurable logic fabric.

Currently, we implement the dynamic partitioning module as another embedded MicroBlaze processor core including separate instruction and data memories, which can either be located on-chip or off-chip depending on what is acceptable for a given warp processor implementation. Alternatively, one could eliminate the need for the DPM by executing the partitioning tools as a software task on the main processor sharing computation and memory resources with the main application.

Figure 3 shows the overall organization of our warp configurable logic architecture. The WCLA consists of a data address generator (DADG) with loop control hardware (LCH), three input and output registers, a 32-bit multiplier-accumulator (MAC), and utilizes the unused configurable logic within the FPGA to implement the partitioned critical regions. The configurable logic architecture handles all memory accesses to and from the configurable logic using the data address generator. Furthermore, the data retrieved and stored to and from each array is located within one of the three registers Reg0, Reg1, and Reg2.

**Figure 3:** Warp configurable logic architecture (WCLA) for dynamic hardware/software partitioning.

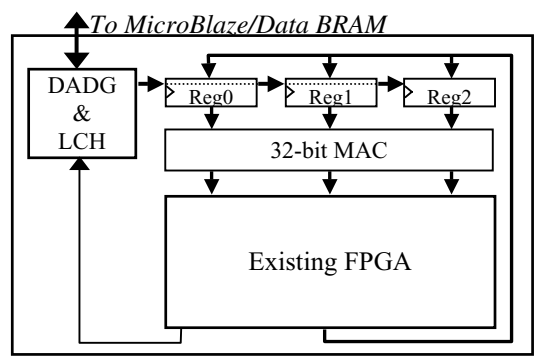

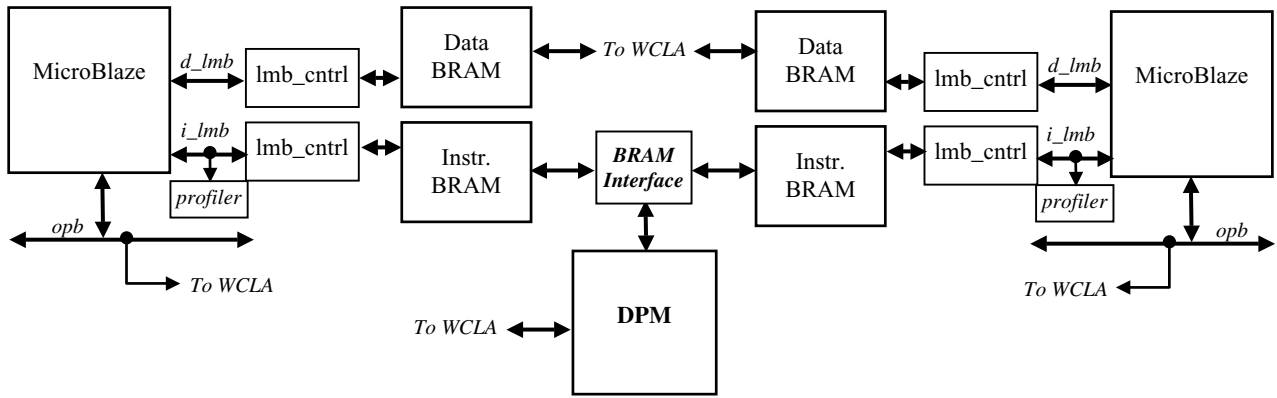

**Figure 4:** MicroBlaze multi-processor warp processing system.

These three registers also act as the inputs to the configurable logic fabric and can be mapped as inputs to the 32-bit (MAC) or directly mapped to the configurable logic fabric. Finally, the WCLA connect the outputs from the configurable logic fabric as inputs to the three registers using a dedicated bus.

Since warp processing targets critical loops that typically iterate many times before completion, the WCLA must be able to access memory and to control the execution of the loop. Therefore, the WCLA includes a DADG with LCH to handle all memory accesses as well as to control the execution of the loop. The DADG within the WCLA can handle memory accesses that follow a regular access patterns. The DADG and LCH interface with the MicroBlaze system by accessing the dual ported data BRAM to read and write data during the hardware execution. The WCLA also communicates with the MicroBlaze processor using the on-chip peripheral bus.

Ideally, our proposed MicroBlaze-based warp processor would utilize the FPGA's unused configurable logic to re-implement the application's critical regions. However, developing computer aided design tools for existing FPGAs capable of executing on-chip using very limited memory resources is a difficult task [16][17]. Instead, our MicroBlaze warp processor utilized a simple configurable logic fabric for re-implementing the critical kernels in hardware that we developed simultaneously with a set of lean synthesis, technology mapping, placement, and routing algorithms [15]. By targeting a simplified configurable logic fabric, our ROCPART tools can execute on a small, embedded processor requiring very little memory and execution time while producing good results. Alternatively, we could create a virtual configurable logic fabric by superimposing our custom configurable logic fabric on top of the underlying physical FPGA fabric, which we are currently investigating.

As mentioned earlier, one the benefits of using a soft processor core is the ability for designers to build systems with multiple processors using the same FPGA. The number of processors a designer can incorporate within any given FPGA is only limited by the size of the FPGA itself. However, in building a multi-processor warp processor device, simply instantiating multiple warp processors as shown in Figure 2 would results in a very large overhead as multiple DPMs and WCLAs are not needed to support a multi-processor system. Instead, a single DPM is sufficient for performing partitioning and synthesis for each of the processors in a round robin or similar fashion. Furthermore, one could again implement the partitioning tools as a software task executing on any of the multiple processors.

Figure 4 presents a multi-processor MicroBlaze warp processor. For each individual MicroBlaze processor we include a profiler. While we could create a profiler capable of monitoring all processors simultaneous, the profiler is very small compared with the processor and should not have significant impact on area or energy consumption. The dynamic partitioning module in a multiprocessor system needs to be able to interface to all processors individually. Therefore, we include a simple *BRAM Interface* that allows the DPM to select which processor's memory to access. Finally, we need to modify our original WCLA to include separate data address generators, loop control hardware, registers, and MAC for each of the processors. However, as the requirements for the applications executing on each of the processors are likely to differ, the critical regions implemented in hardware for the various processors can share the available configurable logic.

## 4. Experiments

In determining the performance and energy benefits of warp processing for the MicroBlaze soft processor core, we analyzed the execution time and power consumption of several embedded systems applications from the Powerstone and EEMBC benchmark suites. Our experimental setup considers a MicroBlaze processor system implemented using the Spartan3 FPGA. When implemented on a Spartan3 FPGA, the MicroBlaze processor core has a maximum clock frequency of 85 MHz. However, the remaining FPGA circuits can operate at up 250 MHz. We configured the processor to include a barrel shifter and multiplier, as the applications we considered required both operations.

Figure 6 and Figure 7 present the speedup and energy reduction of the MicroBlaze-based warp processor compared with a standard MicroBlaze processor. We simulated the software application execution on the MicroBlaze using the Xilinx Microprocessor Debug Engine and obtained an instruction trace for each application. We used the instruction trace to simulate the behavior of the on-chip profiler to determine the single most critical region within each application. Using the profiling results,

**Figure 5:** Equation for determining energy consumption after hardware/software partitioning.

$$E_{total} = E_{MB} + E_{HW} + E_{static}$$
$$E_{MB} = P_{MB(idle)} \times t_{idle} + P_{MB(active)} \times t_{active}$$
$$E_{HW} = P_{HW} \times t_{active}$$
$$E_{static} = P_{static} \times t_{total}$$

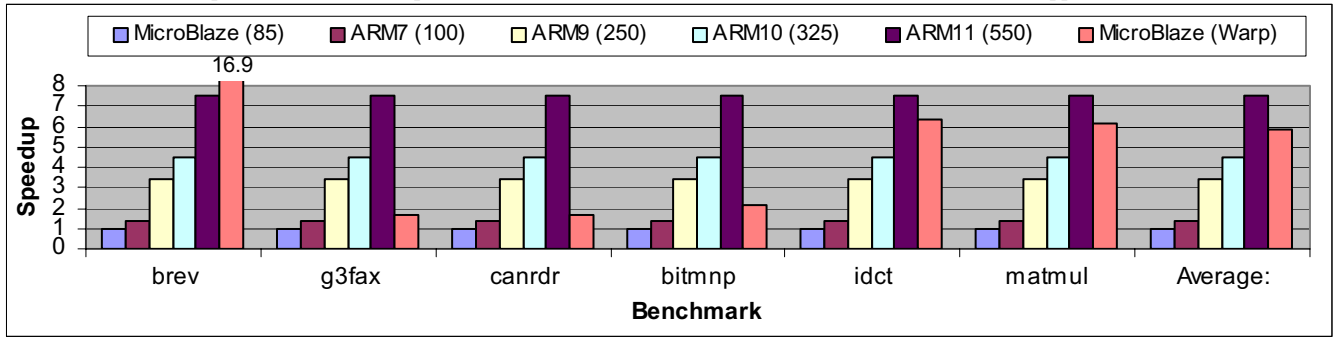

**Figure 6:** Speedups of MicroBlaze-based warp processor and ARM7, ARM9, ARM10, and ARM11 (MHz in parentheses) processors compared to MicroBlaze processor alone for various Powerstone and EEMBC benchmark applications.

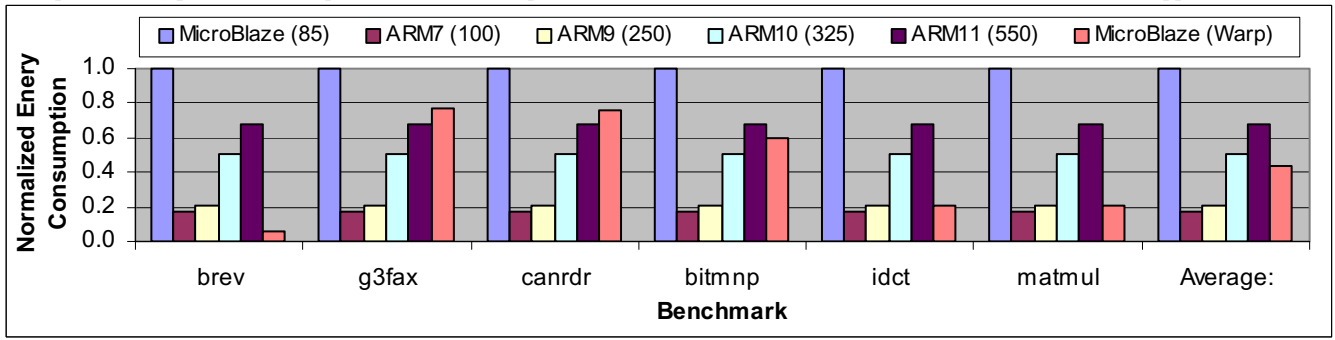

**Figure 7:** Normalized energy consumption of MicroBlaze-based warp processor and ARM7, ARM9, ARM10, and ARM11 (MHz in parenthesis) processors compared to MicroBlaze processor alone for various Powerstone and EEMBC benchmark applications.

we executed our ROCPART partitioning tools for each application to create the hardware circuit implementations targeting our simple configurable logic fabric. To determine the performance and power consumption of the WCLA, we implemented our configurable logic architecture in VHDL and synthesized the design using Synopsys Design Compiler targeting the UMC 0.18 µm technology library. Then, using the execution traces, we simulated the hardware circuits for each partitioned critical region using the VHDL model of the WCLA to determine the final application performance with warp processing. Finally, we calculated the energy consumption of each of the applications using the equations in Figure 5. With the exception of the configurable logic fabric, we determined the power consumption of the MicroBlaze processor and system components using the Xilinx XPower power estimation tool to determine the dynamic and static power consumption. We also compared the performance and energy consumption of the MicroBlaze processor and MicroBlaze warp processor to several ARM processors. For each application, we determined the execution for the ARM processors using the SimpleScalar simulator ported for the ARM processor [4].

Figure 6 presents the speedups of a MicroBlaze-based warp processor and several ARM hard-core processors compared with a MicroBlaze soft processor core. The MicroBlaze warp processor is able to improve the performance of a MicroBlaze system, achieving an average of speedup of 5.8. The largest speedup is for the application *brev*, achieving a speedup 16.9. As we described earlier, *brev*'s critical kernel is a bit reversal implemented using shift operations and other bit manipulations. However, after partitioning the kernel to hardware, the result hardware circuit is much more efficient, requiring only wires to implement the bit reversal. Excluding the application *brev*, which is a special situation in which a hardware implementation is much more efficient than software, the MicroBlaze warp processor achieves an average speedup of 3.6 over a MicroBlaze processor alone. We note that the MicroBlaze warp processor achieves these performance results using the simple configurable logic fabric and not the more robust native configurable logic of the Spartan3 FPGA. Instead, if we could target the native Spartan3 fabric, we would expect to see additional performance improvements.

Figure 7 presents the normalized energy consumption of the MicroBlaze-based warp processor and several ARM hard-core processors compared with a MicroBlaze soft processor core. While having the lowest clock frequency, the MicroBlaze processor has the highest energy consumption, requiring 48% more energy than the ARM11, which has the second highest overall energy consumption. However, the MicroBlaze warp processor is able reduce the energy consumption by an average of 57%, with a maximum energy reduction of 94% for the application *brev*. Excluding *brev*, the MicroBlaze warp processor achieves an average energy reduction of 49%.

While the MicroBlaze warp processor provides increased performance and lower energy consumption compared to the MicroBlaze processor, we also need to compare the MicroBlaze warp processor with readily available hard-core processors. Overall, the MicroBlaze warp processor has better performance than the ARM7, ARM9, and ARM10 processors and requires less energy than the ARM10 and ARM11 processors. The ARM11 processor executing at 550 MHz is on average 2.6X faster than the MicroBlaze warp processor but requires 80% more energy. Furthermore, compared with the ARM10 executing at 325 MHz, the MicroBlaze warp processor is on average 1.3X faster while requiring 26% less energy. Therefore, while the MicroBlaze warp processor is not the fastest nor the lowest energy alternative, the MicroBlaze warp processor is comparable and competitive with existing hard-core processors, while having all the flexibility advantages associated with soft-core processors.

Of course, one could also consider applying warp processing to a single-chip hard-core microprocessor/FPGA device, and in fact that was the original focus our warp processing technique. What we have shown in this paper is that warp processing can make a soft-core FPGA processor competitive with a standalone hard-core processor, which simply expands the usefulness of soft-cores on FPGAs to cover a broader range of applications with tighter performance and power constraints, providing a competitive alternative to a hard-core processor.

## 5. Conclusions

Soft-core processors provided by FPGA vendors give designers the flexibility to configure the processors and enable those designers to quickly build FPGA systems incorporating one or more processors. However, these soft processor cores, such as the MicroBlaze, typically require longer execution times with higher energy consumption compared with hard-core processors, limiting the number of potential applications in which a designer would choose to use a soft processor core. We studied a MicroBlaze-based warp processor that dynamically and transparently optimizes the executing application by re-implementing critical software kernels in hardware using the configurable logic available within the FPGA. The MicroBlaze warp processor eliminates the performance and energy overhead by improving performance on average by 5.8X and reducing energy consumption on average by 57%, making them competitive with current hard-core processors. Designers using MicroBlaze warp processors can quickly build FPGA systems incorporating a MicroBlaze or similar soft processor core with performance and energy consumption comparable to standard hard-core processors such as an ARM processor. With warp processing, soft processors cores can be used in a wider range of application in which using a soft-core processor would not have been previously feasible due to performance and/or energy requirements.

Our current future work includes creating physical implementations of warp processing, improving warp processing's tools to provide speedups for a wider range of applications, and making use of advanced on-chip configurable structures for further improvements [18].

## 6. Acknowledgements

This research was supported in part by the National Science Foundation (CCR-0203829), the Semiconductor Research Corporation (2003-HJ-1046G), and Xilinx Corp.